\documentclass[12pt]{iopart}

\usepackage{graphicx}
\usepackage{color}
\usepackage{bm}
\usepackage{amssymb}
 \usepackage{cite}

\begin{document}

\title[Role of Mutual Information in Entropy Production under Information Exchanges]{Role of Mutual Information in Entropy Production under Information Exchanges}

\author{Takahiro Sagawa$^1$ and Masahito Ueda$^2$}

\address{
${}^1$
Department of Basic Science, The University of Tokyo, 3-8-1 Komaba, Meguro-ku, Tokyo 153-8902, Japan \\
${}^2$ Department of Physics, The University of Tokyo, 7-3-1 Hongo, Bunkyo-ku, Tokyo 113-0033, Japan \\
}
\ead{sagawa@noneq.c.u-tokyo.ac.jp}
\begin{abstract}
We relate the information exchange between two stochastic systems to the nonequilibrium entropy production in the whole system.  By deriving a general formula that decomposes the total entropy production into the thermodynamic and informational parts, we obtain  nonequilibrium equalities such as the fluctuation theorem in the presence of information processing.  Our results apply not only to situations under measurement and feedback control, but also to those under multiple information exchanges between two systems, giving the fundamental energy cost for information processing and elucidating the thermodynamic and informational roles of a memory in information processing.
We clarify a dual relationship between measurement and feedback.
\end{abstract}

\maketitle

\section{Introduction}

Thermodynamics of information processing has seen a resurgence of interest recently.  From a theoretical point of view, the advances in nonequilibrium statistical mechanics over the last two decades have opened up a new avenue of research to generally and quantitatively investigate the relationship between nonequilibrium thermodynamics and information theory~\cite{Nielsen,Touchette,Touchette2,Cao1,Kim,Sagawa-Ueda1,Cao2,Touchette3,Jacobs,Sagawa-Ueda3,Suzuki,Horowitz1,Morikuni,Sagawa,Jennings,SWKim,Ito,Horowitz2,Abreu,Jarzynski5,Pekola,Horowitz3,Abreu3,Sagawa-Ueda4,Munakata,Esposito,Abreu2,Lahiri,Sagawa2,Fetio,Piechocinska,Allahverdyan3,Horhammer,Barkeshli,Norton,Maroney,Turgut,Sagawa-Ueda2,Lutz1,Esposito2,Vedral,Lambson,Still,Sagawa-Ueda2012,Granger,Ito2,Diana,Deffner,Tasaki,Mandal,Barato,Strasberg,HSP,Mandal2,Barato2}, shedding new light on the longstanding problem concerning Maxwell's demon~\cite{
Demon,Maxwell,Szilard,Brillouin,Landauer,Bennett}.
From an experimental point of view,  developments in  experimental techniques have led to the realization of Maxwell's demon with small thermodynamic systems~\cite{Lopez,Toyabe,Berut}.

Furthermore,  the nonequilibrium equalities such as the fluctuation theorem (FT)~\cite{Cohen,Jarzynski1,Crooks1,Crooks2,Jarzynski2,Kurchan,Tasaki0,Jarzynski3,Seifert,Kawai,Jarzynski4,Campisi,Sagawa3} have been generalized to the case under information processing.  For example, we have derived a generalized FT in the presence of an information exchange~\cite{Sagawa-Ueda2012}.  However, a fundamental question remains elusive:  What is the relationship between the exchanged information inside the universe and the total entropy production in the  universe?  Here, the ``universe''  means the relevant entire system including heat baths.

In the present paper, we address this question by focusing on the role of the mutual information in the total entropy production in the whole system.
By deriving a decomposition formula of the total entropy production into the thermodynamic and informational parts, 
we investigate FT and the second law of thermodynamics (SL) in the presence of information processing.
In particular, we examine SL under multiple information exchange.
We also point out that there exists a certain duality between measurement and feedback, which relates the entropic cost for measurement to that for feedback.
Moreover, we study the detailed structure of a memory that stores information, and obtain a general formula that determines the fundamental energy cost needed for measurement and feedback control.

All of our results are based on the detailed fluctuation theorem (DFT)~\cite{Crooks1,Crooks2,Jarzynski2}, and  are therefore  not restricted to the near-equilibrium regime.
Our theory provides the basis for understanding the entropic and energetic properties of information-driven nanomachines~\cite{Mandal,Barato,Strasberg,HSP,Mandal2,Barato2}.

This paper is organized as follows. 
In Sec.~2, we consider the case of a single information exchange, and derive a general formula of the decomposition of the entropy production.
In Sec.~3, we consider the case of multiple information exchanges, and apply the obtained general result to the composite process of measurement and feedback control; this process includes a typical setup of Maxwell's demon.
In Sec.~4, we analyze the entropic and informational roles played by the memory, which enables us to derive the minimal energy cost needed for measurement.
In Sec.~5, we conclude this paper.
In Appendix A, we discuss the entropy production in the heat bath, and clarify the physical meaning of the total entropy production along the line with the standard nonequilibrium statistical mechanics.

\section{Single information exchange}

In this section, we consider the case of a single information exchange.  In Sec.~2.1, we briefly review as much of information theory as is needed for later discussions.
In Sec.~2.2, we derive a general formula of the decomposition of the entropy production under information processing.  In Sec.~2.3 and 2.4, we apply the general formula to situations under feedback and measurement, respectively.  In Sec.~2.5, we discuss a duality between measurement and feedback.

\subsection{Information contents}

We first review the Shannon entropy (or information) and the mutual information~\cite{Shannon,Cover-Thomas}, which play key roles in following discussions.

Let $x$ be a probability variable with probability distribution $P[x]$.  The stochastic Shannon entropy is defined by
\begin{equation}
s[x] := - \ln P[x],
\end{equation}
which characterizes how rare the occurrence of an outcome $x$ is; the rarer it is, the greater $s[x]$ becomes.
The average of $s[x]$ over the probability distribution $P[x]$ gives the Shannon entropy
\begin{equation}
\langle s_x \rangle := - \sum_x P[x] \ln P[x].
\label{Shannon}
\end{equation}
If $x$ is a continuous variable, the sum in Eq.~(\ref{Shannon}) is replaced by the integral.

Let $x$ and $y$ be two probability variables with joint probability distribution $P[x,y]$.  The marginal distributions are given by $P[x] := \sum_y P[x,y]$ and $P[y] := \sum_x P[x,y]$.
The stochastic mutual information is defined by
\begin{equation}
I [x,y] := \ln \frac{P[x,y]}{P[x]P[y]}.
\end{equation}
The ensemble average of $I[x,y]$ gives the mutual information:
\begin{equation}
\langle I \rangle := \sum_{xy} P[x,y] \ln \frac{P[x,y]}{P[x]P[y]}.
\end{equation}
The mutual information characterizes the correlation between the two probability variables.  We also note the relation
\begin{equation}
\langle s_{xy} \rangle = \langle s_x \rangle + \langle s_y \rangle - \langle I \rangle,
\label{mutual_Shannon1}
\end{equation}
where
\begin{equation}
\langle s_y \rangle := - \sum_y P[y] \ln P[y], \ \langle s_{xy} \rangle := - \sum_{xy} P[x,y] \ln P[x,y].
\end{equation}

The Shannon entropy of $x$ and the mutual information between $x$ and $y$ satisfy the following inequalities:
\begin{equation}
0 \leq \langle I \rangle \leq \langle s_x \rangle,
\label{mutual_Shannon2}
\end{equation}
where the left equality is achieved if and only if the two variables are not correlated, or equivalently statistically independent (i.e., $P[x,y] = P[x] P[y]$); the right equality is achieved if and only if, for any $y$, there exists a unique $x$ such that $P[x,y] \neq 0$.  A parallel argument holds true if  we replace  $\langle s_x \rangle$ by $\langle s_y \rangle$ in Eq.~(\ref{mutual_Shannon2}).

\subsection{Decomposition formula}

We consider stochastic dynamics of two systems $X$ and $Y$ in the presence of information exchange between them. 
We assume that  $X$ is attached to  heat baths with inverse temperatures $\beta_k$ ($k = 1, 2, \cdots$).
We denote the baths collectively as $B$. 
System $X$ then evolves under the influence of system $Y$, where we assume that the phase-space point of $Y$ at a particular time, denoted as $y$, only affects the dynamics of $X$ (see also Fig.~1).
We note that the present situation is  the same  as the one in our previous paper~\cite{Sagawa-Ueda2012}, but we here adopt a different approach to deriving FT and SL.

\begin{figure}[htbp]
 \begin{center}
 \includegraphics[width=50mm]{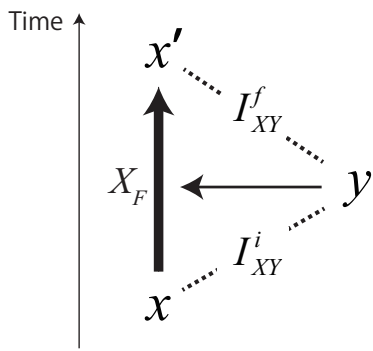}
 \end{center}
 \caption{Time evolution of $X$ under the influence of $Y$.  System $X$ evolves from $x$ to $x'$ along trajectory $X_F$, where the phase-space point of $Y$ at a particular time, denoted as $y$, only affects the dynamics of $X$ .  There may be initial and final correlations between $X$ and $Y$ which are characterized by mutual information contents $I_{XY}^i$ and $I_{XY}^f$.} 
\end{figure}

Let $x$ and $x'$ be the initial and final phase-space points of $X$, and $y$ be the phase-space point of $Y$.
Let $P_F^i [x, y]$ and $P_F^f[x', y]$ be the initial and final joint probability distributions of the composite system $XY$.  
Here the subscript ``$F$'' indicates the ``forward process.''
We define $P_F^i [x] := \int dy P_F^i[x, y]$, $P_F^f [x'] := \int dy P_F^f[x', y]$, and $P_F [y] := \int dx P_F^i[x, y] = \int dx' P_F^f[x', y]$. 
We note that the marginal distribution of $y$ does not change in time.
We assume that there may, in general, be the initial and final correlations between $X$ and $Y$, i.e., $ P_F^i[x, y] \neq P_F^i [x] P_F [y] $ and $ P_F^f[x, y] \neq P_F^f [x] P_F [y]$.

We consider the difference between the Shannon entropy of $(x,y)$ and that of $(x',y)$, which is given by
\begin{equation}
\Delta s_{XY} :=   (- \ln P_F^f [x', y]) - (- \ln P_F^i [x, y]).
\end{equation}
It can be rewritten as 
\begin{equation}
\Delta s_{XY} = \Delta s_X + \Delta s_Y - \Delta I_{XY},
\end{equation}
where 
\begin{equation}
\Delta s_X :=   (- \ln P_F^f [x']) - (- \ln P_F^i [x]), 
\end{equation}
\begin{equation}
\Delta s_Y :=   (- \ln P_F [y]) - (- \ln P_F [y]) = 0, 
\end{equation}
\begin{equation}
\Delta I_{XY} := I_{XY}^f - I_{XY}^i := \ln \frac{P_F^f[x', y]}{P_F^f [x']P_F[y]} - \ln \frac{P_F^i [x, y]}{P_F^i [x] P_F[y]}.
\end{equation}
Since $\Delta s_Y = 0$, we obtain
\begin{equation}
\Delta s_{XY} = \Delta s_X - \Delta I_{XY}.
\end{equation}
In the following, we denote the initial and final Shannon entropies of $X$ as 
\begin{equation}
s_X^i := - \ln P_F^i [x], \ s_X^f := - \ln P_F^f [x'].
\end{equation}

Let $Q_{X,k}$ be the heat absorbed by system $X$ from the $k$th bath.  Following to the standard nonequilibrium thermodynamics~\cite{Crooks2,Jarzynski2,Seifert}, the entropy production in the total system ($X$, $Y$, and $B$) during the present dynamics is given by
\begin{equation}
\Delta s_{XYB} := \Delta s_{XY} + \Delta s_B,
\label{entropy_total}
\end{equation}
where 
\begin{equation}
\Delta s_B := - \sum_k \beta_k Q_{X,k}
\label{entropy_bath}
\end{equation}
is the entropy production in $B$ (see Appendix A for details).
We then obtain the decomposition of the total entropy production as follows:
\begin{equation}
\eqalign{
\Delta s_{XYB} &= \Delta s_{X} - \Delta I_{XY} + \Delta s_B \\
&= \Delta s_{XB} - \Delta I_{XY},
}
\label{main1}
\end{equation}
where
\begin{equation}
\Delta s_{XB} := \Delta s_{X}  + \Delta s_B
\end{equation}
is the entropy increase in $XB$.

We examine the above result in terms of DFT.
Let $X_F$ be the trajectory of $X$ in the forward process.
The joint probability distribution of $X_F$ and $y$ is given by
\begin{equation}
P_F [X_F,y] = P_F [X_F | x, y]P_F^i [x, y],
\end{equation}
where $P_F [X_F | x, y]$ is the conditional probability of $X_F$ under the initial condition $(x,y)$, where the dependence on $y$ reflects the effect of information exchange.  We write the ensemble average of an arbitrary quantity $A[X_F, y]$ as
\begin{equation}
\langle A \rangle := \int dX_F dy P_F [X_F,y] A[X_F, y].
\end{equation} 

To formulate DFT, we need to introduce the concept of backward processes, where the time dependence of external parameters such as the magnetic field is time-reversed.
The backward probability distribution is given by
\begin{equation}
P_B [X_B, \tilde y] = P_F [X_B | \tilde x, \tilde y ]P_B^i [\tilde x, \tilde y ],
\end{equation}
where $P_B [X_B | \tilde x, \tilde y]$ is the conditional probability of $X_B$ under the initial condition $(\tilde x, \tilde y)$. 

Let $x^\ast$ and $y^\ast$ be the time-reversal of the phase-space points $x$ and $y$, respectively.  For example, if $x = (\bm r, \bm p)$ with position $\bm r$ and momentum $\bm p$, then $x^\ast = (\bm r, - \bm p)$.  For $X_F = \{ x(t) \}_{0 \leq t \leq \tau}$, we define its time-reversal as $X_F^\dagger := \{ x^\dagger(t) \}_{0 \leq t \leq \tau} := \{ x^\ast (\tau - t) \}_{0 \leq t \leq \tau}$.  
In a broad class of nonequilibrium dynamics, the entropy production in $B$ satisfies~\cite{Crooks1,Crooks2,Jarzynski2}
\begin{equation}
\Delta s_{B} = \ln \frac{P_F [X_F | x, y ] }{P_B [X_F^\dagger | x'^\ast, y^\ast ]},
\label{DFT0}
\end{equation}
where the left-hand side (lhs) is the entropy production in $B$ in the forward process, and the right-hand side (rhs) is the ratio of the probability distributions of the forward and backward trajectories.
We then assume that the initial distribution of the backward processes is given by the time-reversal of the final distribution of the forward process:
\begin{equation}
P_B^i[x', y] := P_F^f [x'^\ast, y^\ast ],
\end{equation}
which leads to DFT for the total system:
\begin{equation}
\Delta s_{XYB} = \ln \frac{P_F [X_F ,y] }{P_B [X_F^\dagger ,y^\ast]}.
\end{equation}
We then have
\begin{equation}
\eqalign{
\Delta s_{XYB} &= \ln \frac{P_F [X_F | x, y ] }{P_B [X_F^\dagger | x'^\ast, y^\ast ]} + \ln \frac{P_F^i [x, y]}{P_F^f [x', y]} \\
&= \ln \frac{P_F [X_F | x, y ] }{P_B [X_F^\dagger | x'^\ast, y^\ast ]} + \ln \frac{P_F^i [x | y]}{P_F^f [x' |  y]} \\
&= \ln \frac{P_F [X_F | x, y ] }{P_B [X_F^\dagger | x'^\ast, y^\ast ]} + \ln \frac{P_F^i[x]}{P_F^f[x']} +  \ln \frac{P_F^i [x | y]}{P_F^i[x]}  + \ln \frac{P_F^f[x']}{P_F^f [x' |  y]} \\
&= \ln \frac{P_F [X_F | x, y ]P_F^i[x] }{P_B [X_F^\dagger | x'^\ast, y^\ast ]P_F^f[x']} +  \ln \frac{P_F^i [x | y]}{P_F^i[x]}  - \ln \frac{P_F^f [x' |  y]}{P_F^f[x']}.
}
\end{equation}
By noting that 
\begin{equation}
\Delta s_{XB} = \ln \frac{P_F [X_F | x, y ]P_F^i[x] }{P_B [X_F^\dagger | x'^\ast, y^\ast ]P_F^f[x']},
\end{equation}
we reproduce Eq.~(\ref{main1}).

In the present setup, the Kawai-Parrondo-van den Broeck (KPB) equality~\cite{Kawai} is given by
\begin{equation}
\langle \Delta s_{XYB} \rangle = \langle \Delta s_{XB} \rangle - \langle \Delta I_{XY} \rangle = \int dX_F dy  P_F [X_F ,y] \ln \frac{P_F [X_F ,y] }{P_B [X_F^\dagger ,y^\ast]},
\label{KPB1}
\end{equation}
where the rhs is the relative entropy between the forward and backward trajectories.  From the positivity of the relative entropy~\cite{Cover-Thomas}, we obtain SL for the total process:
\begin{equation}
\langle \Delta s_{XYB} \rangle \geq 0,
\label{second1}
\end{equation}
which is equivalent to
\begin{equation}
\langle \Delta s_{XB} \rangle \geq \langle \Delta I_{XY} \rangle.
\label{second2}
\end{equation}
Inequality~(\ref{second2}) implies that the lower bound of the entropy increase in $XB$ is given by the change in the mutual information between $X$ and $Y$.

Let $\mathcal S$ be the set of $(x,y)$ such that $P_F^i [x, y] \neq 0$.  We  then have
\begin{equation}
\langle e^{- \Delta s_{XYB}} \rangle = \int_{\mathcal S} dX_F dy P_F [X_F ,y] \frac{P_B [X_F^\dagger ,y]}{P_F [X_F ,y] } = \int_{\mathcal S} dX_F^\dagger dy^\ast P_B [X_F^\dagger ,y ^\ast],
\end{equation} 
where we used $dX_F = dX_F^\dagger$ and $dy = dy^\ast$.
If $\mathcal S$ is the whole phase space, we obtain the integral fluctuation theorem (IFT) or the Jarzynski equality:
\begin{equation}
\langle e^{- \Delta s_{XYB}} \rangle = 1,
\end{equation}
which is equivalent to
\begin{equation}
\langle e^{- \Delta s_{XB} + \Delta I_{XY}} \rangle = 1.
\label{IFT1}
\end{equation}

The crucial assumption here is that the dynamics of $X$ is affected only by the phase-space point $y$ at a particular time.  Therefore,  $Y$ does not necessarily stay at $y$ as $X$ evolves, as long as the evolution of $Y$ does not affect the dynamics of $X$.  Therefore,  the probability distribution of $X_F$ is characterized by $P_F [X_F | x, y]$ that is not affected by the time evolution of $Y$.

Although we have obtained the same results as (\ref{KPB1}), (\ref{second2}), and IFT~(\ref{IFT1}) in a previous paper~\cite{Sagawa-Ueda2012}, we stress that in this paper we have adopted a new approach to deriving them on the basis of the decomposition formula~(\ref{main1}).
The present approach gives a new insight compared with the previous one, in that it enables us to understand the generalized FT and SL as a result of the decomposition of the total entropy production.
We note that a decomposition formula similar to Eq.~(\ref{main1}) has been discussed in Ref.~\cite{HSP} for special cases.

In the absence of information exchange, $P_F [X_F | x, y]$ is independent of $y$ so that $P_F [X_F | x]$.   In this case, $\Delta s_{XB}$ satisfies the conventional DFT and therefore its expectation value is nonnegative:
\begin{equation}
\langle \Delta s_{XB} \rangle = \int dX_F P_F [X_F ] \ln \frac{P_F [X_F | x ]P_F^i[x] }{P_B [X_F^\dagger | x'^\ast ]P_F^f[x']} \geq 0.
\label{no_information}
\end{equation}
We also have
\begin{equation}
\eqalign{
\langle \Delta I_{XY} \rangle &=  \int dX_F dy  P_F[X_F, y] \ln \frac{P_F^f [x' |  y]P_F^i[x]}{P_F^f[x']P_F^i [x | y]} \\
&=  \int dX_F dy  P_F[X_F, y] \ln \frac{P_F^f [x' |  y] P_F [X_F | x ] P_F^i[x]}{P_F^f[x']P_F [X_F | x ] P_F^i [x | y]} \\
&= \int dX_F dy  P_F[X_F, y] \ln \frac{P_F [X_F | x' ]}{P_F [X_F | x', y ]} \\
&=  - \int dX_F dy P_F^f[x']  P_F[X_F, y | x'] \ln \frac{P_F [X_F, y | x' ]}{P_F [X_F | x' ]P_F[y | x']} \\
&\leq 0,
}
\end{equation}
which is a special case of the data processing inequality~\cite{Cover-Thomas}.
Therefore, in the absence of information processing, we obtain
\begin{equation}
\langle \Delta s_{XYB} \rangle =\langle \Delta s_{XB} \rangle - \langle \Delta I_{XY} \rangle \geq \langle \Delta s_{XB} \rangle \geq 0.
\end{equation}
In other words, inequality~(\ref{no_information}) is stronger than inequality~(\ref{second2}) in this case; $\langle \Delta s_{XB} \rangle$ cannot be negative due to inequality~(\ref{no_information}), even when inequality~(\ref{second2}) gives a negative lower bound.
Therefore, in the absence of information exchange,  it is consistent to regard $XB$ as the whole ``universe'' even when there are initial and final correlations with $Y$; we can ignore what's happening outside $XB$ if there is no interaction between inside and outside of the universe.

\subsection{Feedback control}

We apply the foregoing general framework to feedback control, where  $X$ is the system to be controlled and $Y$ is the memory that initially has the information about the initial condition of  the system and controls it depending on that information  (see also Fig.~2 (a)). The mutual information that is initially shared between the system and the memory is given by $I := I_{XY}^i$, and the final remaining correlation is given by  $I^{\rm rem} := I_{XY}^f$.  The decomposition (\ref{main1}) of the total entropy production is then given by
\begin{equation}
\Delta s_{XYB} = \Delta s_{XB} +(I - I^{\rm rem}),
\label{feedback0}
\end{equation}
which, together with inequality~(\ref{second1}), leads to
\begin{equation}
\langle \Delta s_{XB} \rangle \geq  - \langle I - I^{\rm rem} \rangle. 
\label{feedback1}
\end{equation}
Inequality~(\ref{feedback1}) implies that the entropy in $XB$ can be decreased by the amount up to $\langle I - I^{\rm rem} \rangle$ that characterizes the upper bound of the utilized information during feedback control.

\begin{figure}[htbp]
 \begin{center}
 \includegraphics[width=120mm]{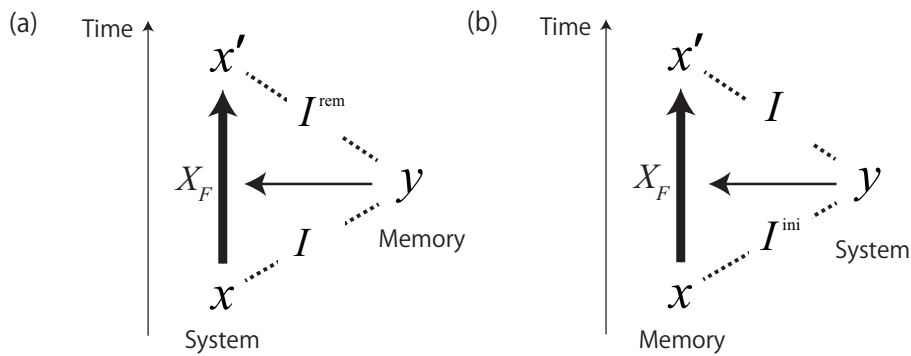}
 \end{center}
 \caption{ (a) Dynamics of feedback control, where $X$ is the system to be controlled and $Y$ is the memory.   (b) Dynamics of measurement, where $X$ is the memory and $Y$ is the measured system.  These schematics illustrate the dual relationship between measurement and feedback control; they have a one-to-one correspondence under time-reversal and exchange of the roles of the system and the memory.} 
\end{figure}

We next consider the energetics of feedback control.
Let $E^i_{X} [x ]$ and   $E^f_{X,y} [x' ]$ be the initial and final Hamiltonians of system $X$.  
Here, we assume that the initial Hamiltonian is independent of $y$, and that the final one can depend on $y$ through feedback control.  
The intermediate Hamiltonians during the feedback process can also depend on $y$.
For simplicity, we neglect the interaction Hamiltonian between $X$ and $Y$ in the initial and final states.
The energy change in this process is given by
\begin{equation}
\Delta E_{X} := E^f_{X,y} [x' ] - E^i_{X} [x].
\end{equation}
The first law of thermodynamics is given by
\begin{equation}
\Delta E_{X} = \sum_k Q_{X,k} + W_X,
\end{equation}
where $W_X$ is the work performed on $X$ through the time dependence of external parameters.

We now assume that there is a single heat bath at inverse temperature $\beta$.  Inequality~(\ref{feedback1}) then reduces to
\begin{equation}
\langle W_X \rangle \geq \langle \Delta F_{\rm eff} \rangle  - \beta^{-1} \langle I - I^{\rm rem} \rangle,
\end{equation}
where $\Delta F_{\rm eff} $ is the change in the effective (nonequilibrium) free energy defined by
\begin{equation}
\Delta F_{\rm eff} := \Delta E_{X} - \beta^{-1} \Delta s_X.
\end{equation}
We next define the initial and final equilibrium free energies as follows:
\begin{equation}
F^i_{X} := -\beta^{-1} \ln  \int dx e^{-\beta E^i_X [x]}, \ F^f_{X,y} := -\beta^{-1} \ln \int dx' e^{-\beta E^f_{X,y} [x' ]}.
\end{equation}
We further assume that the initial distribution of $X$ is the thermal equilibrium:
\begin{equation}
P^i_F [x ] = e^{\beta (F^i_{X} - E_{X}^i [x])}.
\end{equation}
We then obtain
\begin{equation}
\langle E_{X}^i - \beta^{-1} s_X^i \rangle = F_{X}^i.
\end{equation}
On the other hand, the final distribution can be different from the canonical distribution in general.  Let $\tilde s^f_{X,y} [x' ] := - \ln P_F^f [x' | y ]$ be the conditional Shannon entropy of the final distribution.  We then have an inequality:
\begin{equation}
\langle E^f_X - \beta^{-1} \tilde s^f_{X,y}  \rangle \geq F^f_{X,y},
\end{equation}
where the equality is achieved if and only if $P_X^f [x' | y ]$ is the conditional canonical distribution for a given $y$:
\begin{equation}
P^f_F [x' | y] = e^{\beta (F^f_{X,y} - E_{X,y}^f [x'])}.
\end{equation}
We note that $- \ln P_F^f [x'] = - \ln P_F^f [x'|y] + \ln ( P_F^f [x'|y] / P_F^f[x'] )$, and therefore
\begin{equation}
s^f_X = \tilde s^f_{X,y} + I^{\rm rem}.
\end{equation}
We finally obtain
\begin{equation}
\langle W_X \rangle \geq \langle \Delta F_X \rangle  - \beta^{-1} \langle I \rangle,
\label{feedback2}
\end{equation}
where
\begin{equation}
\langle \Delta F_X \rangle := \sum_y P_F[y] F^f_{X,y} - F_X^i
\end{equation}
is the average change in the conditional free energy.
Inequality~(\ref{feedback2}) sets the fundamental lower bound of the energy cost for feedback control, which is smaller by the amount of $\beta^{-1} \langle I \rangle$ than the usual thermodynamic bound.  We note that the same bound as~(\ref{feedback2}) has been obtained in Refs.~\cite{Sagawa-Ueda1,Sagawa-Ueda3} for a different setup.

\subsection{Measurement}

We next apply our general framework to measurement processes, where $X$ is the memory and $Y$ is the measured system  (see also Fig.~2 (b)).
In other words, $X$ performs a measurement on $Y$ in this setup.
We first assume that the initial correlation is zero (i.e., $I_{XY}^i = 0$) before the measurement, and the final correlation is characterized by the information ($I := I_{XY}^f$) obtained by the measurement.
The total entropy production is given by
\begin{equation}
\Delta s_{XYB} = \Delta s_{XB} -I,
\end{equation}
which, together with inequality~(\ref{second1}), leads to
\begin{equation}
\langle \Delta s_{XB} \rangle \geq \langle I \rangle.
\label{meas1}
\end{equation}
Inequality~(\ref{meas1}) implies that the entropy in $XB$ inevitably increases due to the obtained information by the measurement.

If the memory has prior knowledge about the system before the measurement, there is the corresponding initial correlation $I_{\rm ini} := I_{XY}^i$.  We then obtain
\begin{equation}
\Delta s_{XYB} = \Delta s_{XB} - (I - I^{\rm ini}),
\end{equation}
which, together with inequality~(\ref{second1}), leads to
\begin{equation}
\langle \Delta s_{XB} \rangle \geq \langle I  - I^{\rm ini} \rangle.
\label{meas2}
\end{equation}
Inequality~(\ref{meas2}) implies that the entropy increase in $XB$ is bounded from below by the obtained information $\langle I  - I^{\rm ini} \rangle$.

To discuss the energetics of the memory, we need to examine the more detailed structure of the memory, which will be discussed in Sec.~4.

\subsection{Duality between measurement and feedback control}

We now discuss a fundamental relationship between measurement and feedback control.
Let us consider the time-reversal transformation of the dynamics and exchange the roles  of the system and the memory at the same time (see also Fig.~2).
We then find that the measurement becomes feedback and vice versa, where $I$ in measurement corresponds to $I$ in feedback, and $I^{\rm ini}$ in measurement corresponds to $I^{\rm rem}$ in feedback.
This implies a kind of dual structure between the measurement and feedback, as summarized in Table 1.

\begin{table}
\caption{Duality between measurement and feedback.}
\begin{tabular}{| l || l | l |} \hline
{} & Measurement & Feedback \\ \hline \hline
Role of $X$ & Memory & System \\ \hline 
Role of $Y$ & System & Memory \\ \hline 
Initial correlation & $I^{\rm ini}$ &  $I$ \\ \hline
Final correlation & $I$ &  $I^{\rm rem}$ \\ \hline
Second law & $\langle \Delta s_{XB} \rangle \geq \langle I - I^{\rm ini}\rangle$ & $\langle \Delta s_{XB} \rangle \geq \langle I^{\rm rem} - I \rangle$ \\ \hline 
\end{tabular}
\end{table}

We consider a special case of $\langle I^{\rm ini} \rangle  = \langle I^{\rm rem} \rangle = 0$.
In this case, the lower bound of  $\langle \Delta s_{XB} \rangle$ is given by $\langle I \rangle$ for measurement and by $-\langle I \rangle$ for feedback, where the opposite signs are due to the fact that   the final correlation in measurement corresponds to the initial correlation in feedback because of the time-reversal transformation.
This explains the reason why the entropy in $XB$ is increased by measurement but decreased by feedback control.


\section{Multiple information exchanges}

We generally consider the case of multiple information exchanges in Sec.~3.1, and then focus on the case of Maxwell's demon in Sec.~3.2.

\subsection{General framework}

We consider multiple information exchanges between two systems $X$ and $Y$, which are attached to different heat baths with each other. 
For simplicity, we use notation $B$ to indicate all baths.  
If the correlation time in the baths is sufficiently small compared with the time scale of the systems, we may apply this assumption to the situation in which the systems are attached to the same baths.
We consider a composite process consisting of the following two processes (see also Fig.~3 (a)). 

\begin{figure}[htbp]
 \begin{center}
 \includegraphics[width=120mm]{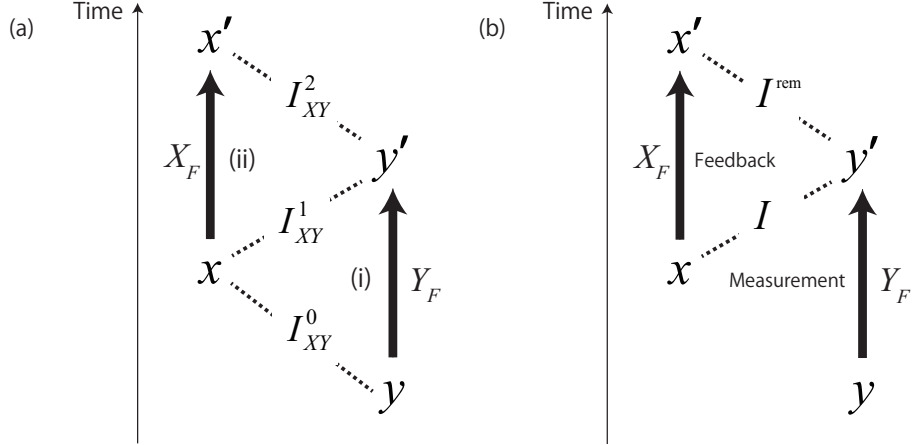}
 \end{center}
 \caption{(a) Dynamics of the two-step composite process.  In the first process (i), $Y$ evolves under the influence of the initial phase-space point of $X$, denoted by $x$.   In the second process (ii), $X$ evolves under the influence of the final phase-space point of $Y$, denoted by  $y'$.   (b) Typical situation of Maxwell's demon. $X$ is the system to be controlled and $Y$ is the memory of the demon, where the first process describes the measurement with outcome $y'$ and the second process describes the feedback control.} 
\end{figure}

In the first process (i), $Y$ evolves under the influence of the initial phase-space point of  $X$, denoted as $x$.
Let $P_F^0 [x,y]$ be the initial distribution of the first process.  System $Y$ evolves along trajectory $Y_F$ with probability $P_F [Y_F |  x, y ]$ under the initial condition of $(x,y)$.  The final distribution of $Y$ is given by $P_F^1 [x,y']$, where $y'$ is the final phase-space point of $Y$.  Let $\Delta s_{XYB}^{\rm (i)}$ and $\Delta s_{YB}^{\rm (i)}$ respectively be the entropy productions in $XYB$ and $YB$ in this process.
The change in the mutual information is given by 
\begin{equation}
\Delta I_{XY}^{\rm (i)} := I_{XY}^1 - I_{XY}^0 := \ln \frac{P_F^1[x, y']}{P_F^1 [x] P_F^1[y']}  - \ln \frac{P_F^0[x, y]}{P_F^0 [x] P_F^0[y]}.
\end{equation}

In the second process (ii), $X$  evolves under the influence of the final phase-space point of $Y$, denoted as $y'$ (see Fig.~3 (a)).
Let $P_F^1 [x,y']$ be the initial distribution of the second process.  System $X$ evolves along trajectory $X_F$ with probability $P_F [X_F |  x, y' ]$ under the condition of $(x,y')$.  The final distribution of $X$ is given by $P_F^2 [x',y']$, where $x'$ is the final phase-space point of $X$.  Let $\Delta s_{XYB}^{\rm (ii)}$ and $\Delta s_{XB}^{\rm (ii)}$ be the entropy productions in $XYB$ and $XB$ in this process.
The change in the mutual information is given by 
\begin{equation}
\Delta I_{XY}^{\rm (ii)}  := I_{XY}^2 - I_{XY}^1 := \ln \frac{P_F^2[x', y']}{P_F^2 [x'] P_F^2[y']}  - \ln \frac{P_F^1[x, y']}{P_F^1 [x] P_F^1[y']}.
\end{equation}

The total entropy production in the composite process, denoted by $\Delta s_{XYB}^{\rm tot}$, is given by the sum of the entropy productions of the two processes:
\begin{equation}
\eqalign{
\Delta s_{XYB}^{\rm tot} &=  \Delta s_{XYB}^{\rm (i)} + \Delta s_{XYB}^{\rm (ii)} \\
&= ( \Delta s_{XB}^{\rm (i)} - \Delta I_{XY}^{\rm (i)}) + ( \Delta s_{YB}^{\rm (ii)}  - \Delta I_{XY}^{\rm (ii)}) \\
&= \Delta s_{XB}^{\rm (i)} + \Delta s_{YB}^{\rm (ii)}  - \Delta I_{XY}^{\rm tot}.
}
\label{composite1}
\end{equation}
The change in the mutual information in the total process is given by
\begin{equation}
\Delta I_{XY}^{\rm tot} = \ln \frac{P_F^2[x', y']}{P_F^2 [x'] P_F^2[y']} - \ln \frac{P_F^0[x, y]}{P_F^0 [x] P_F^0[y]},
\end{equation}
which can also be expressed as  the sum of the changes in the two processes:
\begin{equation}
\Delta I_{XY}^{\rm tot} = \Delta I_{XY}^{\rm (i)} + \Delta I_{XY}^{\rm (ii)}.
\end{equation}

In terms of DFT,  the entropy productions are given by
\begin{equation}
\Delta s_{XYB}^{\rm (i)} = \ln \frac{P_F [Y_F |  x, y ]P_F^0[x, y]}{P_B [Y_F^\dagger | x^\ast, y'^\ast]P_F^1 [x, y' ]}, \ \Delta s_{YB}^{\rm (i)} = \ln \frac{P_F [Y_F |  x, y ]P_F^0[ y]}{P_B [Y_F^\dagger | x^\ast , y'^\ast ]P_F^1 [y']},
\end{equation}
\begin{equation}
\Delta s_{XYB}^{\rm (ii)} = \ln \frac{P_F [X_F |  x, y' ]P_F^1[x, y']}{P_B [X_F^\dagger | x'^\ast, y'^\ast]P_F^2 [x', y']}, \ \Delta s_{XB}^{\rm (ii)} = \ln \frac{P_F [X_F |  x^\ast, y'^\ast ]P_F^1[ x]}{P_B [X_F^\dagger | x'^\ast , y'^\ast  ]P_F^2 [x' ]},
\end{equation}
and
\begin{equation}
\Delta s_{XYB}^{\rm tot} = \ln \frac{P_F [X_F |  x, y' ]P_F [Y_F |  x, y ]P_F^0[x, y]}{P_B [Y_F^\dagger | x^\ast, y'^\ast]P_B [X_F^\dagger | x'^\ast, y'^\ast]P_F^2 [x', y']}.
\end{equation}
Here, we have assumed that the initial distributions of the two backward processes are given by $P_F^1 [x^\ast, y'^\ast]$ and $P_F^2 [x'^\ast, y'^\ast]$.  

We note that the initial distribution of the backward process  of (i) is not necessarily equal  to the final distribution of the backward process of  (ii).  In other words, the first backward process is not necessarily followed by the second backward process; one cannot start the backward process of (i) immediately after the backward process of (ii), but one should change the probability distribution to start the backward process of (ii).
On the other hand, the initial distribution of the forward process  (i) is  equal  to the final distribution of the forward process (ii).  Therefore, the forward process (i) is actually followed by the forward process (ii), and one can start the forward process (ii) immediately after the forward process (i).  

Since the total entropy production is nonnegative, we obtain
\begin{equation}
\langle \Delta s_{XYB}^{\rm tot} \rangle \geq 0,
\end{equation}
and therefore
\begin{equation}
\langle \Delta s_{XB}^{\rm (i)} \rangle + \langle \Delta s_{YB}^{\rm (ii)}  \rangle \geq \langle \Delta I_{XY}^{\rm tot} \rangle.
\label{main_c}
\end{equation}
Inequality (\ref{main_c}) implies that the sum of the entropy increases is bounded by the total change in the mutual information.

We note that the foregoing argument can straightforwardly be generalized to the case of information exchanges which take place more than once.

\subsection{Maxwell's demon}

We next consider the composite process of measurement and feedback, which is a typical situation of Maxwell's demon (see also Fig.~3 (b)).  
In this case, $X$ is the system to be controlled and $Y$ is the memory of the demon.
We assume that there is no initial correlation: $I_{XY}^{0} = 0$.  After the measurement, the memory obtains the mutual information $I_{XY} := I_{XY}^{1}$ and then uses it for feedback control.  The remaining correlation after feedback control is given by $I^{\rm rem}_{XY} := I_{XY}^{2}$. 
By applying Eq.~(\ref{composite1}) to this case, the total entropy production of the composite process is given by
\begin{equation}
\eqalign{
\Delta s_{XYB}^{\rm tot} &=  \Delta s_{XYB}^{\rm meas} + \Delta s_{XYB}^{\rm feed} \\
&=  ( \Delta s_{YB}^{\rm meas} - I_{XY} )+ ( \Delta s_{XB}^{\rm feed} +(I_{XY}-I_{XY}^{\rm rem})) \\
&= \Delta s_{XB}^{\rm feed} + \Delta s_{YB}^{\rm meas} -  I_{XY}^{\rm rem}.
}
\end{equation}
Therefore, we obtain
\begin{equation}
\langle \Delta s_{XB}^{\rm feed} \rangle + \langle \Delta s_{YB}^{\rm meas} \rangle \geq \langle  I_{XY}^{\rm rem} \rangle.
\end{equation}
Since $\langle I_{XY}^{\rm rem} \rangle$ is non-negative, we obtain
\begin{equation}
\langle \Delta s_{XB}^{\rm feed} \rangle + \langle \Delta s_{YB}^{\rm meas} \rangle \geq 0.
\end{equation}
This inequality implies that the entropy decrease in $XB$ by feedback control is compensated for by the entropy increase in $YB$ by measurement.

We note that, the total entropy productions $\langle \Delta s_{XYB}^{\rm meas} \rangle$ and $\langle \Delta s_{XYB}^{\rm feed} \rangle$ are both nonnegative during measurement and feedback, which confirms that the role of the demon does not contradict SL.  The crucial observation here is that the mutual information $ \langle I_{XY} \rangle$ which is stored during the measurement is used as a resource of the entropy decrease during the feedback process.

\section{Memory structure}

We next discuss the detailed structure of the memory, and its roles in  measurement and feedback control.

\subsection{Setup and decomposition of entropy}

We consider a situation in which the phase space of the memory, which we refer to as $\mathcal Y$, is divided into several  subspaces (see also Fig.~4).  Each subspace is written as $\mathcal Y_m$ labeled by $m$ ($= 1,2, \cdots$), where $M := \{ m \}$ may be regarded as the set of measurement outcomes.  
We assume that $\mathcal Y_m$'s do not overlap with each other, and $\bigcup_m \mathcal Y_m = \mathcal Y$. For any $y \in \mathcal Y$, there is a single $m$ such that $y \in \mathcal Y_m$, which we write as $m_y$.

\begin{figure}[htbp]
 \begin{center}
 \includegraphics[width=100mm]{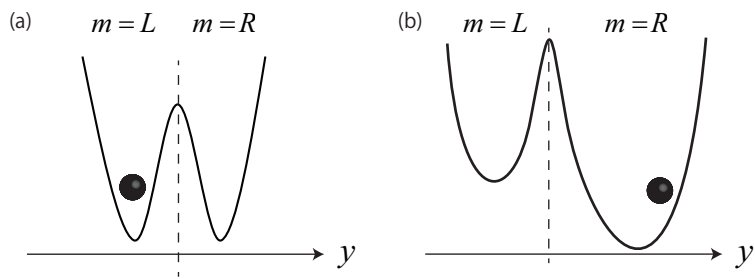}
 \end{center}
 \caption{Schematic of the double-well memory structure with $m= L, R$.   (a) Symmetric memory with $F^f_{Y,L} = F^f_{Y,R}$.  (b) Asymmetric memory $F^f_{Y,L} \neq F^f_{Y,R}$.} 
\end{figure}

We consider probability distribution $P[y]$ over $\mathcal Y$.
Let $p[m]$ be the probability of $y \in \mathcal Y_m$, and $P [y|m]$ be the conditional probability of $y$ under the condition of $y \in \mathcal Y_m$.  We note that  $P[y|m] = 0$  if $m \neq m_y$, because $\mathcal Y_m$'s do not overlap with each other. The joint probability distribution is given by
\begin{equation}
P[y, m] = P[y|m]p[m] \delta (m, m_y),
\end{equation} 
where $\delta (\cdot, \cdot)$ is the Kronecker delta.
The unconditional probability distribution is then given by
\begin{equation}
P[y] = \sum_m P[y, m]  =  P[y|m_y]p[m_y].
\end{equation}
We define the stochastic Shannon entropies as
\begin{eqnarray}
s_Y[y] &:=& -\ln P[y], \\
s_{Y,m} [y] &:=& -\ln P[y|m], \\
h_M[m] &:=& - \ln p[m],
\end{eqnarray}
which satisfy
\begin{equation}
s_Y[y] =  h_M[m_y]+ s_{Y,m_y} [y].
\end{equation}
Therefore, we obtain
\begin{equation}
\langle s_Y \rangle = \langle h_M \rangle + \langle \tilde s_Y \rangle, 
\label{prob_formula}
\end{equation}
where
\begin{equation}
\langle s_Y \rangle = - \int_{\mathcal Y} dy P[y] \ln P [y],
\end{equation}
\begin{equation}
\langle \tilde s_Y \rangle := - \sum_{m} p[m] \langle s_{Y, m} \rangle = - \sum_{m} \int_{\mathcal Y_{m}} dy P[y|m] p[m] \ln P[y|m],
\end{equation}
\begin{equation}
\langle h_M \rangle  = - \sum_m p[m] \ln p[m].
\end{equation}
Equality~(\ref{prob_formula}) implies that the total Shannon entropy is decomposed into the Shannon entropy over $m$ and the average Shannon entropy of the phase-space points in $\mathcal Y_m$, where the former characterizes the randomness of the measurement outcomes, while the latter characterizes the average of the fluctuations within individual subspaces.



\subsection{Measurement}

We now consider measurement processes with the memory structure in the presence of heat baths $B$.
Let us choose a subspace  $\mathcal Y_0$ which may be one of $\mathcal Y_m$'s, but not necessarily be so.
In fact, $\mathcal Y_0$ may be equal to the whole phase space $\mathcal Y$.
We assume that the initial phase-space point $y$ is in $\mathcal Y_0$ with unit probability; in this case, we say that the memory is in the standard state. 
Let $ P_F^{i} [y] $ be the initial distribution of $y$; by assumption, $P^i_F[y]=0$ if $y$ does not belong to $\mathcal Y_0$.
 We also assume that there is no initial correlation between $X$ and $Y$.
 
The memory then evolves along trajectory $Y_F$ under the influence of $X$ with phase-space point $x$, and stores outcome $m$ with probability $p_F[m]$.
This measurement establishes the correlation between $x$ and $m$.
After the measurement,  the final phase-space point is $y'$.  We note that the probability that $y'$ is in subspace  $\mathcal Y_m$ is given by $p_F[m]$.
Let $P_F^{f} [y'|m]$ be the final probability distribution of $y'$ under the condition of $m$.

The total entropy production during the measurement is then given by
\begin{equation}
\Delta s_{XYB}^{\rm meas} =   h_M + \Delta \tilde s_Y^{\rm meas}  + \Delta  s_B^{\rm meas}  -   I_{XY},
\end{equation}
where
\begin{equation}
\Delta s_{XYB}^{\rm meas} = \ln \frac{P_F [Y_F, x] }{P_B [Y_F^\dagger, x^\ast]},
\end{equation}
\begin{equation}
h_M := - \ln p_F [m],
\end{equation}
\begin{equation}
\Delta \tilde s_Y^{\rm meas} := (- \ln P_F^{f} [y'|m]) - ( - \ln P_F^{i} [y] ),
\end{equation}
\begin{equation}
I_{XY} := \ln \frac{P_F [x,y']}{P_F[x]P_F[y']}.
\end{equation}
In the following, we write  $ s_{Y}^{i} := - \ln P_F^{i} [y]$ and $ s_{Y, m}^{f} := - \ln P_F^{f} [y'|m]$.

We next assume that there is a single heat bath at inverse temperature $\beta$. 
 Let $E_{Y,0}^i [y]$ be the initial Hamiltonian defined on subspace $\mathcal Y_0$.  We assume that the initial distribution is given by the canonical distribution in $\mathcal Y_0$:
\begin{equation}
P_F^i [y] = e^{\beta (F_{Y,0}^i - E_{Y, 0}^i [y])},
\end{equation}
where the conditional free energy is given by
\begin{equation}
F_{Y,0 }^i := - \beta^{-1} \ln \int_{\mathcal Y_0} dy  e^{-\beta  E_{Y,0}^i [y]}.
\end{equation}
In this case,
\begin{equation}
F_{Y,0}^i  = \langle E_{Y,0}^{i} - \beta^{-1} s_{Y}^{i} \rangle.
\end{equation}
Let  $E_{Y,m}^f [y']$  be the  final Hamiltonian defined only on $\mathcal Y_m$.  We define the conditional free energy as
\begin{equation}
F^{f}_{Y, m} := - \beta^{-1} \ln \int_{\mathcal Y_m} dy'  e^{-\beta  E^{f}_{Y,m} [y']}.
\end{equation}
We refer to the memory as symmetric if $F^{f}_{Y, m}$ takes on the same value for all $m$ (see also Fig.~4).
We then have
\begin{equation}
F_{Y,m}^{f}  \leq \langle E_{Y,m}^{f} - \beta^{-1} s_{Y, m}^{f} \rangle,
\end{equation}
where the equality is achieved if and only if
\begin{equation}
P_F^{f} [y' |m ] = e^{\beta (F_{Y,m}^{f} - E_{Y,m}^{f} [y'])},
\end{equation}
which vanishes outside of $\mathcal Y_m$.
We then have
\begin{equation}
\langle  \Delta E_Y^{\rm meas}  -  \beta^{-1} \Delta \tilde s_Y^{\rm meas}  \rangle \geq \langle \Delta F_Y^{\rm meas} \rangle,
\end{equation}
where
\begin{equation}
\Delta E_Y^{\rm meas} := E_{Y,m}^{f} [y'] - E_{Y,0}^{i} [y],
\end{equation}
\begin{equation}
\langle \Delta F_Y^{\rm meas} \rangle := \sum_m p[m] F_{Y,m}^f - F_{Y,0}^i.
\end{equation}
Therefore, we have
\begin{equation}
\langle \Delta s_{XYB}^{\rm meas} \rangle \geq \beta \langle W_Y^{\rm meas} \rangle - \beta \langle\Delta F_Y^{\rm meas} \rangle + \langle  h_M \rangle - \langle   I_{XY} \rangle,
\end{equation}
where $W_Y^{\rm meas}$ is the work performed on the memory during the measurement. 
Since $\langle \Delta s_{XYB}^{\rm meas} \rangle \geq 0$, we finally obtain
\begin{equation}
\langle W_Y^{\rm meas} \rangle \geq    \langle\Delta F_Y^{\rm meas} \rangle - \beta^{-1} \langle  h_M \rangle + \beta^{-1} \langle   I_{XY} \rangle,
\label{meas_e1}
\end{equation}
which determines the minimal energy cost for measurement. 
The lower bound is characterized by the average free-energy difference, the Shannon information of measurement outcomes, and the mutual information between $X$ and $Y$.
On the rhs of inequality~(\ref{meas_e1}), $- \beta^{-1} \langle  h_M \rangle$ arises from the increase in the Shannon entropy of the memory by the measurement, and $\beta^{-1} \langle   I_{XY} \rangle$ arises from the increase of the mutual information between the system and the memory by the measurement.
The reason why the signs of  $- \beta^{-1} \langle  h_M \rangle$ and  $\beta^{-1} \langle   I_{XY} \rangle$ are different from each other is that the Shannon information and the mutual information contribute to the total entropy with opposite signs as shown in Eq.~(\ref{mutual_Shannon1}).

We note that the actually utilizable information obtained by the memory is characterized by the mutual information between $X$ and outcome $M$:
\begin{equation}
I_{XM} := \ln \frac{P_F^f[x,m]}{P_F[x]p_F[m]},
\end{equation}
where $P_F^f[x,m]$ is the joint probability distribution of $x$ and $m$ after the measurement.
We then have
\begin{equation}
\eqalign{
I_{XY} - I_{XM} &= \ln \frac{P_F^f[x,y]p_F[m]}{P_F^f[x,m]P_F^f[y]} =  \ln \frac{P_F^f[x|y]}{P_F^f[x|m]} \\
&= \ln \frac{P_F^f[x|y,m]}{P_F^f[x|m]} =: \tilde I_{XY},
}
\label{conditional}
\end{equation}
where $P_F^f[x|y]$ and $P_F^f[x|m]$ are the conditional probabilities of $x$ under the condition of $y$ and $m$, respectively.  The ensemble average $\langle  \tilde I_{XY} \rangle$ is the conditional mutual information between $X$ and $Y$ under the condition of $m$, which is by construction nonnegative [see Eq.~(\ref{conditional})]:
\begin{equation}
\langle I_{XY} \rangle - \langle I_{XM} \rangle = \langle \tilde I_{XY} \rangle \geq 0. 
\end{equation}
Therefore, we obtain an inequality which is weaker than (\ref{meas_e1}):
\begin{equation}
\langle W_Y^{\rm meas} \rangle \geq   \langle\Delta F_Y^{\rm meas} \rangle -\beta^{-1}  \langle  h_M \rangle + \beta^{-1} \langle   I_{XM} \rangle.
\label{measurement_cost2}
\end{equation}
Inequality~(\ref{measurement_cost2}) is physically more transparent than inequality~(\ref{meas_e1}), because  the lower bound in (\ref{measurement_cost2})  is characterized by the physically utilizable information $\langle   I_{XM} \rangle$ rather than the total correlation $\langle   I_{XY} \rangle$.
We note that the same bound as (\ref{measurement_cost2}) has been derived in Ref.~\cite{Sagawa-Ueda2} for a different setup.

\begin{figure}[htbp]
 \begin{center}
 \includegraphics[width=100mm]{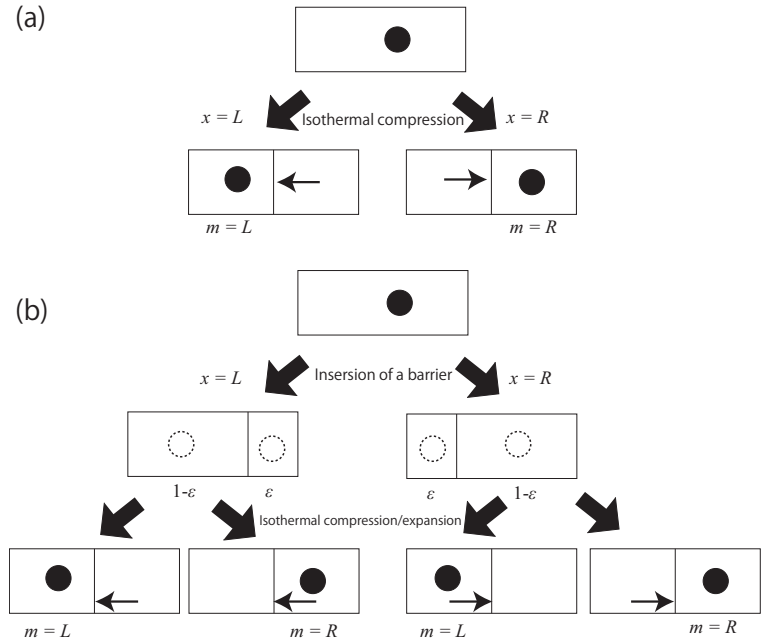}
 \end{center}
 \caption{Simple models of measurement.  (a) Error-free measurement.  The memory is initially in the standard state, which is the global equilibrium in the box.  If the measured state is $x=L$ ($x=R$), the box is compressed from the right (left) quasi-statically and isothermally with the particle confined  in the left (right) box corresponding to $m=L$ ($m=R$).  In the final state, $x$ and $m$ are perfectly correlated.  (b) Measurement with error rate $\varepsilon$.  The standard state is the same as in (a).  If the measured state is $x=L$ ($x=R$),  a barrier is inserted  and the box is divided into two compartments with volume ratio $1-\varepsilon : \varepsilon$ ($\varepsilon : 1-\varepsilon$).  The barrier is moved to the center of the box.  The particle is finally in the left (right) box corresponding to $m=L$ ($m=R$), where $x$ and $m$ are not perfectly correlated if $0 < \varepsilon < 1$.  If $\varepsilon = 0$, this model is equivalent to the error-free model of (a). } 
\end{figure}

As an illustration, we consider a simple model of measurement.
Figure 5 (a) shows a model of  error-free measurement.  The memory is a single particle in a box with a single heat bath at inverse temperature $\beta^{-1}$, where $\mathcal Y_0$ is the whole phase space.
We assume that the measured state is $x=L$ or $R$ with equal probability $1/2$.
After the quasi-static and isothermal measurement described in Fig.~5 (a), the particle is in the left box or the right box corresponding to  $m=L$ or $R$, where  $\mathcal Y_L$ and $\mathcal Y_R$ correspond to the left and right box, respectively.
We note that $x=m$ in this model.
 In this case, $\langle \Delta F_Y^{\rm meas} \rangle = \beta^{-1} \ln 2$, $\langle W_Y^{\rm meas} \rangle = \beta^{-1} \ln 2$, $\langle  h_M \rangle = \ln 2$, and $\langle I_{XM} \rangle = \ln 2$.  Therefore, the equality in inequality (\ref{measurement_cost2}) is achieved in this model.

Figure 5 (b) shows a model of   measurement with error rate $\varepsilon$ ($0 \leq \varepsilon \leq 1$), where $\mathcal Y_0$, $\mathcal Y_L$, and $\mathcal Y_R$ are the same as in the previous example.  
We assume that the measured state is $x=L$ or $R$ with the equal probability of $1/2$.  In this case, $\langle \Delta F_Y^{\rm meas} \rangle = \beta^{-1} \ln 2$, $\langle  h_M \rangle = \ln 2$, and
\begin{equation}
\langle W_Y^{\rm meas} \rangle = \beta^{-1} [ \ln 2 + \varepsilon \ln \varepsilon + (1-\varepsilon ) \ln ( 1- \varepsilon) ],
\end{equation}
\begin{equation}
\langle I_{XM} \rangle =  \ln 2 + \varepsilon \ln \varepsilon + (1-\varepsilon ) \ln ( 1- \varepsilon).
\end{equation}  
Therefore, the equality in  (\ref{measurement_cost2}) is again achieved in this model.

We now briefly discuss the information erasure from the memory.  During the erasure, memory $Y$ is detached from the measured system $X$, and $Y$ returns to the standard state; after the erasure, the phase-space point of $Y$ is in $\mathcal Y_0$ with unit probability.
The Shannon entropy in $M$ after the erasure is $0$ by definition; it changes by $- \langle  h_M \rangle$ during the erasure, whose sign is opposite to that in the measurement.  
Since $Y$ is detached from $X$ during the erasure, DFT and SL can apply to $YB$ (see also arguments in the last paragraph of  Sec.~2.2).
Therefore, the entropy change in $YB$ during the erasure satisfies
\begin{equation}
\langle \Delta s_{YB}^{\rm eras} \rangle \geq 0,
\end{equation}
where the equality can be achieved in the quasi-static erasure.

We assume that there is a single heat bath at inverse temperature $\beta$, and that the probability distribution of $Y$ in each $\mathcal Y_m$ before the erasure is the canonical distribution under the condition of $m$.
By applying a similar argument used in deriving~(\ref{meas_e1}) to $\langle \Delta s_{YB}^{\rm eras} \rangle$, we obtain the lower bound of the  work performed on the memory during the erasure:
\begin{equation}
\langle W_Y^{\rm eras} \rangle \geq    \langle\Delta F_Y^{\rm erase} \rangle + \beta^{-1} \langle  h_M \rangle,
\label{eras}
\end{equation}
which is the generalized Landauer principle~\cite{Maroney,Sagawa-Ueda2}. 
We note that the free-energy change $\langle \Delta F_Y^{\rm eras} \rangle$  during the erasure satisfies  $\langle \Delta F_Y^{\rm eras}  \rangle = - \langle\Delta F_Y^{\rm meas} \rangle$.  
In the special case of $\langle\Delta F_Y^{\rm erase} \rangle  = 0$, inequality~(\ref{eras}) reduces to the conventional Landauer principle~\cite{Landauer,Piechocinska,Berut}, which is satisfied in the case of a symmetric memory as shown in Fig.~4 (a).

By summing up inequalities (\ref{meas_e1}) and (\ref{eras}), the total work for measurement and erasure is  given by
\begin{equation}
\langle W_Y^{\rm meas } \rangle + \langle W_Y^{\rm eras} \rangle \geq \beta^{-1} \langle   I_{XY} \rangle,
\label{trade_off}
\end{equation}
where the lower bound is only determined by the mutual information;  $\langle\Delta F_Y^{\rm meas} \rangle$ and  $- \beta^{-1} \langle  h_M \rangle$ on the rhs of inequality~(\ref{meas_e1}) are canceled by the corresponding terms in inequality~(\ref{eras}).
In fact, the measurement and erasure are  time-reversal with each other if we only focus on $YB$ and ignore the interaction with  $X$. 
However, they are not completely time-reversal if we take into consideration their interaction; $Y$ interacts with $X$ and establishes the correlation only in the measurement process. 
Therefore, the mutual information obtained by the measurement process plays an essential role in determining the work for the entire process of measurement and erasure.

We note that the assumption of the conditional canonical distribution before the erasure is not necessary to derive only inequality~(\ref{trade_off}); we only need to assume that the probability distribution before the erasure is the same as that after the measurement.  In fact, by summing up the entropy changes in measurement and erasure, we obtain $\langle \Delta s_{YB}^{\rm meas} \rangle + \langle \Delta s_{YB}^{\rm eras} \rangle \geq \langle I_{XY} \rangle$.  By applying a similar argument used in deriving~(\ref{meas_e1}) to the entire entropy change $\langle \Delta s_{YB}^{\rm meas} \rangle + \langle \Delta s_{YB}^{\rm eras} \rangle$ in measurement and erasure, we again obtain inequality~(\ref{trade_off}).

\subsection{Feedback control}

We next consider feedback control on $X$ by $Y$  after the measurement.  More precisely, we assume that the dynamics of $X$ is determined only by the outcome $m$. Therefore, we can consider  a composite system $XM$ instead of $XY$.
We assume that system $X$ is attached to heat baths that are different from those in contact with the memory. We denote the baths attached to $X$  again by $B$.

The probability distribution of the forward trajectory of $X$ and $m$ is given by
\begin{equation}
P_F [X_F,m] = P_F [X_F | x, m]P_F^f [x, m],
\end{equation}
where $P_F^f [x, m]$ is the pre-feedback (post-measurement) distribution of $(x,m)$, and $P_F [X_F | x, m]$ is the conditional probability of $X_F$ under the initial condition $(x,m)$ of the feedback process.

The argument is then completely parallel to that in Sec.~2.3 if we replace $Y$ with $M$.
The total entropy production in $XMB$ is given by
\begin{equation}
\Delta s_{XMB}^{\rm feed} = \Delta s_{XB}^{\rm feed} + (I_{XM} - I_{XM}^{\rm rem}),
\end{equation}
where $I_{XM}^{\rm rem}$ describes the remaining correlation after the feedback control.
SL is then expressed as
\begin{equation}
\langle \Delta s_{XB}^{\rm feed} \rangle \geq  - \langle I_{XM} - I_{XM}^{\rm rem} \rangle.
\end{equation}

If there is a single heat bath at inverse temperature $\beta$ and the initial state of system $X$ is in the canonical distribution, we obtain
\begin{equation}
\langle W_X^{\rm feed} \rangle \geq \langle \Delta F_X^{\rm feed} \rangle  - \beta^{-1} \langle I_{XM}  \rangle.
\label{feedback_energy2}
\end{equation}
On the other hand, by considering $XYB$, we can also obtain 
\begin{equation}
\langle W_X^{\rm feed} \rangle \geq \langle \Delta F_X^{\rm feed} \rangle  - \beta^{-1} \langle   I_{XY}   \rangle.
\label{feedback_energy3}
\end{equation}
We note that inequality~(\ref{feedback_energy2}) is stronger than inequality~(\ref{feedback_energy3}) in the present setup.

\section{Conclusion}

We have established the general relationship between the total entropy production of the whole system and the mutual information that is exchanged between two stochastic systems.

In Sec.~2, we have derived the general decomposition formula~(\ref{main1}) for a single information exchange.   Correspondingly, we have obtained the KPB equality~(\ref{KPB1}), SL~(\ref{second2}), and IFT~(\ref{IFT1}), such that they explicitly include the mutual information.  We have applied the general formula to the cases of feedback control (\ref{feedback0}) and measurement~(\ref{meas2}).  In Sec.~3, we have discussed the case of multiple information exchanges, and obtained a general decomposition formula~(\ref{composite1}) and the corresponding SL (\ref{main_c}).   In Sec.~4, we have considered the structure of the memory; its phase space is divided into several subspaces corresponding to the measurement outcomes.  This formulation has clarified the role of the Shannon information of measurement outcomes as well as the mutual information, as shown for the cases of measurement~(\ref{measurement_cost2}) and feedback control (\ref{feedback_energy2}).

Our theory has clarified the role of mutual information in nonequilibrium thermodynamics with information processing, which is not restricted to the conventional case of Maxwell's demon.  As a consequence, we have revealed the  fundamental relationship between the entropy production in the whole universe (system and bath) and the exchanged information inside the universe.  Our results would serve as the theoretical foundation of nonequilibrium thermodynamics of complex systems in the presence of information processing.

\appendix

\section{Entropy production in heat baths}

We consider the entropy change in $B$ in the setup in Sec.~2.  Following the standard approach in nonequilibrium statistical mechanics, we assume that the total system including the baths obeys the Liouville dynamics that conserves the phase-space volume~\cite{Jarzynski1,Jarzynski2,Kawai}.
We also assume that there is no initial correlation between the system and the baths, and that the initial distribution of each bath is given by the canonical distribution~\cite{Jarzynski2,Sagawa3,Esposito3}.

Let $z_k$ be the initial phase-space point of the $k$th bath with $z := (z_1, z_2, \cdots)$,  $E_{B, k} [z_k]$ be the Hamiltonian of the $k$th bath, and $F_{B,k}$ be its free energy.  The initial distribution is given by
\begin{equation}
P_F^i [z] = \prod_k e^{\beta_k (F_{B,k} - E_{B,k}[z_k])} =: P_{\rm can}[z].
\end{equation}
Let $z_k'$ be the final phase-space point of the $k$th bath with $z' := (z_1', z_2', \cdots)$, and $P_B^f [z']$ be the final probability distribution that is in general different from the canonical distribution.   The heat absorbed by the system from the $k$th bath is given by
\begin{equation}
Q_{X,k} := E_{B,k}[z_k] - E_{B,k}[z_k'].
\end{equation}
On the basis of the above definitions along with Eqs.~(\ref{entropy_total}) and (\ref{entropy_bath}), DFT~(\ref{DFT0}) has been shown to hold~\cite{Jarzynski2} even in the presence of the final correlation between the system and the baths.

We note that  $\langle \Delta s_{SYB} \rangle$ is in general different from the change in the Shannon entropy of the total system, while $\langle \Delta s_{SYB} \rangle$ is related to the relative entropy as follows~\cite{Tasaki0,Kawai,Jarzynski4,Sagawa3,Esposito3}:
\begin{equation}
\langle \Delta s_{SYB} \rangle = \int dx' dy dz' P_F^f[x',y,z'] \ln \frac{P_F^f[x',y,z']}{P_F^f[x',y]P_{\rm can}[z']},
\end{equation}
where $P_F^f[x',y,z']$ and $P_F^f[x',y]$ are respectively the final probability distribution of $(x', y, z')$ and $(x', y)$ in the forward process.    Therefore, there are two origins of the positive entropy production in the whole universe: the final correlation between the system and the baths~\cite{Esposito3}, and the lag between the canonical distribution and the final probability distribution of the baths~\cite{Jarzynski4}.
We note that the role of the initial correlation between the system and the baths has been discussed in Refs.~\cite{Allahverdyan3,Horhammer,Jarzynski3,Campisi}.

\ack

We are grateful to Hal Tasaki for valuable discussions. This work was supported by JSPS KAKENHI Grant
Nos. 25800217 and  22340114,  and  by Platform for Dynamic Approaches to Living System from MEXT, Japan.

\end{document}